\documentclass[pra,aps,preprint,showpacs]{revtex4}
\usepackage{epsfig,amssymb,amsfonts}
\usepackage{epsfig}

\def\opone{\leavevmode\hbox{\small1\kern-3.8pt\normalsize1}}

\begin{document}
\title{Entanglement for a bimodal cavity field interacting with a two-level atom}
\author{Jia Liu, Zi-Yu Chen, Shen-Ping Bu, Guo-Feng Zhang\footnote{Corresponding author. Email:
gf1978zhang@buaa.edu.cn; Phone: 86-10-82338289;
Fax:86-10-82338289}} \affiliation{ Department of Physics, Beijing
University of Aeronautics and Astronautics, Beijing 100083,
China.}

\begin{abstract}
Negativity has been adopted to investigate the entanglement in a
system composed of a two-level atom and a two-mode cavity field.
Effects of Kerr-like medium and the number of photon inside the
cavity on the entanglement are studied. Our results show that
atomic initial state must be superposed, so that the two cavity
field modes can be entangled, moreover, we also conclude that the
number of photon in the two cavity mode should be equal. The
interaction between modes, namely, the Kerr effect, has a
significant negative contribution. Note that the atom frequency
and the cavity frequency have an indistinguishable effect, so a
corresponding approximation has been made in this article. These
results may be useful for quantum information in optics systems.

Keywords: Negativity; Kerr-like medium; Jaynes-Cummings model.
\end{abstract}

\maketitle

Quantum computation, one of the most fascinating applications of
quantum mechanics, has the potential to outperform their classical
counterparts in solving hard problems using much less time. There
has been an ongoing effort to search for various physical systems
that maybe propitious to implement quantum computation. Several
prospective approaches for scalable quantum computation have been
identified \cite{1,2,3,4}. Compared to other physical systems, the
optic quantum can be easily realized in experiments. In quantum
optics, the Jaynes-Cummings (JC) model is one of the exactly
solvable models describing the interaction between a single-mode
radiation field and a two-level atom. It has been realized
experimentally in 1987 \cite{5}. There are many ongoing
experimental and theoretical investigations on the various
extensions of the JC model, such as a bimodal cavity field
\cite{6,7}, two atoms \cite{8,9}, multilevel atoms \cite{10,11},
and so on. A two-level atom interacting with a two-mode cavity
field is discussed here.

In view of the resource character of the entanglement, more
attention has been paid to its quantification, such as the
concurrence, the negativity, the relative entropy of entanglement
etc. Entanglement between two qubits in arbitrary state has been
quantified by concurrence \cite{12,13,14}. It is generally
considered that the two-atomic Wehrl entropy \cite{15} can be used
to quantify the entanglement in the JC model when these modes are
initially prepared in the maximally entangled states \cite{16,17}.
Here we use negativity as the measure and deal with the mixed
state entanglement \cite{18}. Many efforts have been put on the
study of the two-mode JC model, but Kerr effect \cite{19,20,21,22}
has not been considered, and this is the main motivation of the
present paper. The scaled units are used in this work. The
interaction between the field and atom are considered in an ideal
and closed cavity, namely, the field damping and the radioactive
damping \cite{23} are ignored.

The system we considered here is an effective two-level atom with
upper and lower states denoted by $|\uparrow>$ and $|\downarrow>$,
respectively. The corresponding frequencies are $\omega_{a}$ and
$\omega_{b}$, moreover, we denote $\omega_{\alpha}$ as the
transition frequency between states $|\uparrow>$ and
$|\downarrow>$. In the two-photon processes, some intermediate
states $|i>$, i=c, d,$\cdots$ are involved, which are assumed to
be coupled to $|\uparrow>$ and $|\downarrow>$ by dipole-allowed
transition. Let $\omega_{i}$ denote the corresponding frequency of
the atomic energy level $|i>$. There are two requirements:
firstly, the atom interacts with the two cavity fields with
frequencies $\omega_{1}$ and $\omega_{2}$, where $\omega_{1}$+
$\omega_{2}\cong \omega_{\alpha}$; secondly,
$\omega_{a}-\omega_{i}$ and $\omega_{b}-\omega_{i}$ are off
resonance of the one-photon linewidth with $\omega_{1}$ and
$\omega_{2}$. If both are satisfied, then the intermediate states
can be adiabatically eliminated \cite{24} and the effective
Hamiltonian of the system can be written in the rotating-wave
approximation as \cite{25,26}
\begin{eqnarray}
H=\sum_{j=1}^2\omega_ja_j^\dag a_j+\omega_\alpha
\frac{\sigma_z}{2}+\chi(a_1^{\dag2}a_1^{2}+a_2^{\dag2}a_2^{2})+\lambda(a_1a_2\sigma_++a_1^{\dag}a_2^{\dag}\sigma_-),
\end{eqnarray}
\noindent where $a_j^{\dag}(a_j)$ and $\omega_j$ denote the creation
(annihilation) operator and frequency in the $j$th mode, the natural
unit $\hbar=1$ is used throughout the paper. $\sigma_{\pm}=(\sigma_x
\pm i\sigma_y)/2$ is the raising and lowering operators, with
$\sigma_m (m=x,y,z)$ is common Pauli spin operator. $\lambda$ is the
coupling constant between the atom and the modes, known as the Rabi
frequency. $\chi$ is the dispersive part of the third-order
nonlinearity of the Kerr-like medium. Through out the investigation,
we consider that $\omega_1+\omega_2=\omega_\alpha$ (i.e. the
resonance case). After some study, it is found that the frequency of
two modes have the same effect on negativity, so for simplicity, it
is given that $\omega_{1}=\omega_{2}=\omega$ in the subsequent
calculations.

We assume that the initial state of the system has the form of
\begin{eqnarray}
|\Psi(0)\rangle=\cos\theta|n_1,n_2,\uparrow\rangle+\sin\theta|n_1,n_2,\downarrow\rangle,
\end{eqnarray}
\noindent where $n_1$ and $n_2$ are field quantum state in the
Fock representation. Here different values of $\theta$ describe
the states with different amplitudes. In view of the initial
condition and Schr\"{o}dinger equation, the wave function of the
system at time $t$ can be obtained as
\begin{eqnarray}
&&
|\Psi(t)\rangle=a(t)|n_1,n_2,\uparrow\rangle+b(t)|n_1,n_2,\downarrow\rangle\nonumber\\
&&\ \ \ \ \ \ \ \ \
+c(t)|n_1+1,n_2+1,\downarrow\rangle+d(t)|n_1-1,n_2-1,\uparrow\rangle,
\end{eqnarray}
\noindent under the condition
$|a(t)|^2+|b(t)|^2+|c(t)|^2+|d(t)|^2=1$, and
\begin{eqnarray}
&&
a(t)=\frac{1}{2\xi_1}\{e^{-T(i\gamma_1+\xi_1)}\cos\theta [i(e^{2T\xi_1}-1)\zeta(n_1+n_2)+\xi_1(e^{2T\xi_1}+1)]\};\nonumber\\
&&
b(t)=\frac{1}{2\xi_2}\{e^{-T(i\gamma_2+\xi_2)}\sin\theta [-i(e^{2T\xi_2}-1)\zeta(n_1+n_2-2)+\xi_2(e^{2T\xi_2}+1)]\};\nonumber\\
&&
c(t)=-\frac{1}{2\xi_1}[ie^{-T(i\gamma_1+\xi_1)}\eta\cos\theta \sqrt{(1+n_1)(1+n_2)}];\nonumber\\
&&
d(t)=-\frac{1}{2\xi_2}[ie^{-T(i\gamma_2+\xi_2)}\eta\sin\theta (e^{2T\xi_2}-1)\sqrt{n_1n_2}];\nonumber\\
&&
\xi_1=\sqrt{-\eta^2-\zeta^2n_1^2-n_2(\eta^2+\zeta^2n_2^2)-n_1[\eta^2(1+n_2)+2\zeta^2n_2]};\nonumber\\
&&
\xi_2=\sqrt{-\zeta^2n_1^2-\zeta^2(n_2-2)^2+n_1(4\zeta^2-n_2\eta^2-2\zeta^2n_2)};\nonumber\\
&&
\gamma_1=1+n_1+n_2+\zeta(n_1^2+n_2^2);\nonumber\\
&& \gamma_2=-1+2\zeta+(1-2\zeta)n_1+\zeta
n_1^2+n_2+\zeta(n_2-2)n_2,
\end{eqnarray}
\noindent where $\eta=\lambda/\omega$, $\zeta=\chi/\omega$, and
$T=\omega t$. From the above equations, the state density operator
at time $t$, $\rho(t)=|\Psi(t)\rangle\langle\Psi(t)|$, can be
easily derived.

If we know density matrix $\rho_{12}$ of a composite system
composed of subsystem $1$ and $2$, the reduced density operator
for subsystem $1$ is $\rho_{1}=Tr_{2}(\rho_{12})$. In our case,
the density operator of two modes for a given atom state is
$\rho_f(t)=Tr_a\rho(t)$ which can be found in the basis
$\{|i,j\rangle,$ $i=n_{1}-1,$ $n_{1},$ $n_{1}+1$ and $j=n_{2}-1,$
$n_{2},$ $n_{2}+1\}$ as

\begin{eqnarray}
\rho_f(t) = \left({{\begin{array}{*{20}c} \ \ \ {|d|^2}\ \ \hfill
& {0}\ \ \hfill & {0}\ \ \hfill & {0}\ \ \ \ \ \ \ \ \hfill &
{da^*}\ \ \hfill & {0}\ \ \hfill & {0}\ \ \hfill & {0}\ \ \hfill &
{0}

\hfill \\
\ \ \ {0}\ \ \hfill & {0}\ \ \hfill & {0}\ \ \hfill & {0}\ \ \ \ \
\ \ \ \hfill & {0}\ \ \hfill & {0}\ \ \hfill & {0}\ \ \hfill &
{0}\ \ \hfill & {0}
\hfill \\
\ \ \ {0}\ \ \hfill & {0}\ \ \hfill & {0}\ \ \hfill & {0}\ \
\hfill & {0}\ \ \hfill & {0}\ \ \hfill & {0}\ \ \hfill & {0}\ \
\hfill & {0}
\hfill \\
\ \ \ {0}\ \ \hfill & {0}\ \ \hfill & {0}\ \ \hfill & {0}\ \
\hfill & {0}\ \ \hfill & {0}\ \ \hfill & {0}\ \ \hfill & {0}\ \
\hfill & {0}
\hfill \\
\ \ \ {ad^*}\ \ \hfill & {0}\ \ \hfill & {0}\ \ \hfill & {0}
\hfill & {|a|^2+|b|^2}\ \ \hfill & {0}\ \ \hfill & {0}\ \ \hfill &
{0}\ \ \hfill & {bc^*}
\hfill \\
\ \ \ {0}\ \ \hfill & {0}\ \ \hfill & {0}\ \ \hfill & {0}\ \
\hfill & {0}\ \ \hfill & {0}\ \ \hfill & {0}\ \ \hfill & {0}\ \
\hfill & {0}
\hfill \\
\ \ \ {0}\ \ \hfill & {0}\ \ \hfill & {0}\ \ \hfill & {0}\ \
\hfill & {0}\ \ \hfill & {0}\ \ \hfill & {0}\ \ \hfill & {0}\ \
\hfill & {0}
\hfill \\
\ \ \ {0}\ \ \hfill & {0}\ \ \hfill & {0}\ \ \hfill & {0}\ \
\hfill & {0}\ \ \hfill & {0}\ \ \hfill & {0}\ \ \hfill & {0}\ \
\hfill & {0}
\hfill \\
\ \ \ {0}\ \ \hfill & {0}\ \ \hfill & {0}\ \ \hfill & {0}\ \
\hfill & {cb^*}\ \ \hfill & {0}\ \ \hfill & {0}\ \ \hfill & {0}\ \
\hfill & {|c|^2}
\hfill \\

\end{array} }} \right),
\end{eqnarray}
which, generally speaking, is a mixed state. So we introduce
negativity which is the usual measurement of entanglement for a
mixed state and is defined as
\begin{eqnarray}
N(\rho)\equiv\frac{\parallel \rho^{T_A}\parallel_1-1}{2},
\end{eqnarray}
where $\rho^{T_A}$ denotes the partial transpose of $\rho$ with
respect to part $A$, and the trace norm of $\rho^{T_A}$ is equal
to the sum of the absolute values of the eigenvalues of
$\rho^{T_A}$. $N(\rho)>0$ corresponds to a entangled state and
$N(\rho)=0$ corresponds to a separate one.

We perform the diagonalization on density matrix. From the
calculation, one can find that the two cavity mode frequencies
have a indistinguishable effect on the entanglement, and then the
frequencies are supposed to be the same as mentioned above. When
the angle $\theta$ is set to be zero, after a straightforward
calculation it is found that the negativity is zero. This
indicates that the two-level atom initial state is not a
superposition state, and then the two-mode cavity field states
will not be entangled during the time evolution process. The atom
initial state has an important influence on the production of the
entangled cavity mode state. Numerical results of entanglement
measure are presented in Figure 1 to 4.

Figure 1 shows the negativity as a function of time $T$. A
coherent superposition state is chosen as atomic initial state.
Fig.1(a) is the case that the cavity is in a two-mode vacuum
state. Negativity evolves with a period and the maximum value is
0.5 since the cavity systems are two qutrits. Changing the vacuum
state to a more general state, Fig. 1(b) shows a novel feature.
Negativity changes non-smoothly and the period is obviously
smaller than that in Fig.1(a). The maximum is 0.4, does not reach
0.5. It can be easily understood since the noise of the system
exists. When one mode is in a vacuum state and the other is in a
non-zero photon state, the period and the amplitude in Fig. 1(c)
decreases sharply compared with those of Figs. 1(a) and 1(b). This
means that we must prepare almost the same photon number in the
two cavity modes in order to get a higher entanglement and longer
entanglement time. Comparing the figures in Fig.1, the vacuum
state is more useful for the production of entanglement between
the cavity modes.

In Fig. 2 (a-c), negativity is shown as a function of the coupling
constant between the atom and the modes $\eta$, and Kerr medium is
set as $\zeta=10$. In Fig. 2(a), the negativity vibrates
periodically with a maximum amplitude 0.5 when the cavities are in
vacuum states. In Fig. 2(b), the negativity first climbs from zero
nearly linearly, and then increases quaveringly. In this case, two
modes have the same photon number and  are indistinguishable. For
$\eta=0$, the atom does not interact with cavity modes, and the
negativity is zero. Then one can conclude that the coupling
strength between atom and modes decreases the classical noise
effect, then the entanglement between the two-mode cavity field is
enhanced accordingly. But it isn't superior to the case in figure
2(a), in which the classical noise is absent. When one cavity is
vacuum while the other is a common state, the entanglement
increases slowly with the increasing coupling constant, and
finally reaches a maximum value 0.5. These results can be seen
from Fig. 2(c). So, when two cavity modes have the same state,
especially the vacuum state, negativity can easily achieve the
ideal value.

Negativity is plotted in Fig. 3 as a function of Kerr medium
coefficient $\zeta$ and the angle $\theta$. When the Rabi frequency
equals to the transition frequency($\eta=1$), although the number of
photon in two cavity modes is the same, the maximum value can not
reach 0.5. The negativity fluctuates periodically with $\theta$.
When $\theta=\frac{n\pi}{4}$(n is odd), negativity has a maximum
value. Coupling strength between two modes field have a negative
contribution on entanglement. So the Kerr medium effect must be
controlled in order to obtain anticipative entanglement.

In order to clearly show the effects of the initial state and
evolve time on the entanglement producing, negativity as a
function of angle $\theta$ and time $T$ is plotted in Fig. 4. It
is found that negativity is periodic with a period  $\pi/2$ for
the different cavity field state. In Figs. 4(a) and 4(b), two
cavity modes are indistinguishable. When Kerr medium effect
increases, entanglement between the two cavity modes becomes weak,
so that the maximum of negativity can not reach 0.5 in Fig. 4(a).
Figure 4(b) shows that some small wave peaks emerge between two
maximal peaks when the two cavity modes are both in states with
same photon number, which can be attributed to the noise since the
modes are not vacuum. This figure also supports the conclusion
that the cavities' photon number affect the entanglement between
two cavity modes. When one mode changes to be vacuum, negativity
decreases as shown in Fig. 4(c). We also should note that when the
coefficient absolute value of the atomic initial superposition
state equals to each other, the greatest mode entanglement can be
obtained.

In conclusion, we have investigated the entanglement between two
cavity modes. It shows that the entanglement is sensitive to the
atom initial state. A superposition initial state of the atom is
necessary to obtain an entangled state between the cavity modes.
Also, the number of photon inside the cavity, Kerr medium and the
coupling strength between cavity and atom all can greatly affect
the cavity modes entanglement.

\textbf{Acknowledgements}

This work was supported by the National Natural Science Foundation
of China (Grant No.10604053 , 2006CB932603, and 90305026) and
BeiHang Lantian Project.

\newpage
\section*{Caption}

Fig.1: Negativity as a function of time $T$, when $\theta=\pi/4$,
$\eta=1$, $\zeta=0$. (a) $n_1=n_2=0$; (b)
 $n_1=n_2=100$; (c) $n_1=0$, $n_2=100$.
\vspace{5mm}

Fig.2: Negativity as a function of the coupling constant between
the atom and the cavity  $\eta$, when $T=1$, $\theta=\pi/4$,
$\zeta=10$. (a) $n_1=n_2=0$;  (b) $n_1=n_2=100$; (c) $n_1=0$,
$n_2=100$. \vspace{5mm}

 Fig.3: Surface plot of negativity as a function of Kerr medium $\zeta$ and phase angle
$\theta$, when $T=1$, $\eta=1$, $n_1=n_2=100$. \vspace{5mm}

Fig.4: Surface plot of negativity as a function of time $T$ and
phase angle $\theta$, when $\eta=1$, $\zeta=10$. (a) $n_1=n_2=0$;
(b) $n_1=n_2=100$; (c) $n_1=0$, $n_2=100$.

\newpage
\begin{figure}
\epsfig{figure=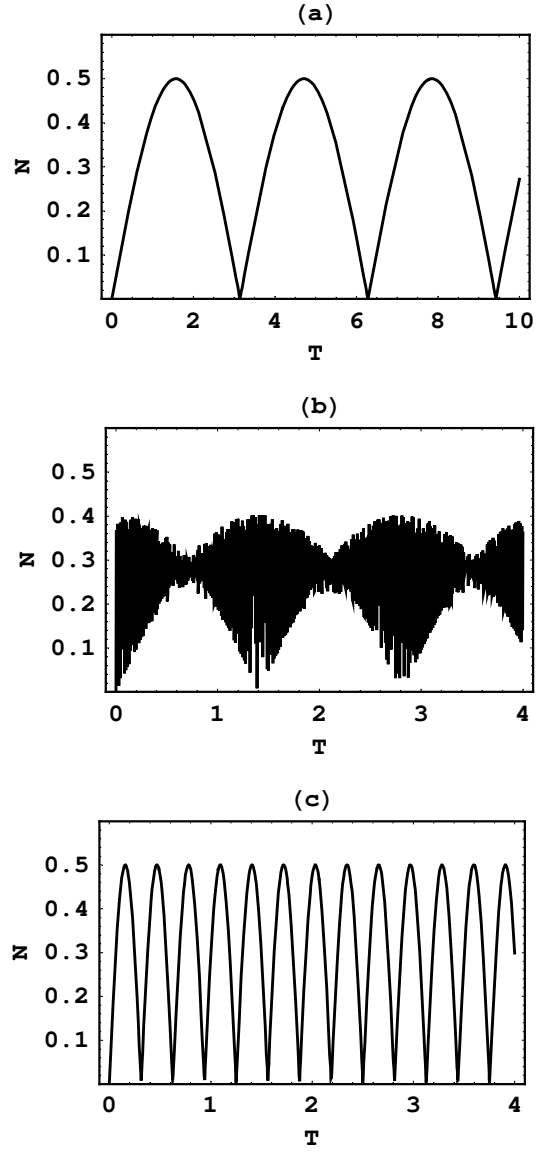} \caption{Negativity as a function of time
$T$, when $\theta=\pi/4$, $\eta=1$, $\zeta=0$. (a) $n_1=n_2=0$;
(b)
 $n_1=n_2=100$; (c) $n_1=0$, $n_2=100$.}
\end{figure}

\begin{figure}
\epsfig{figure=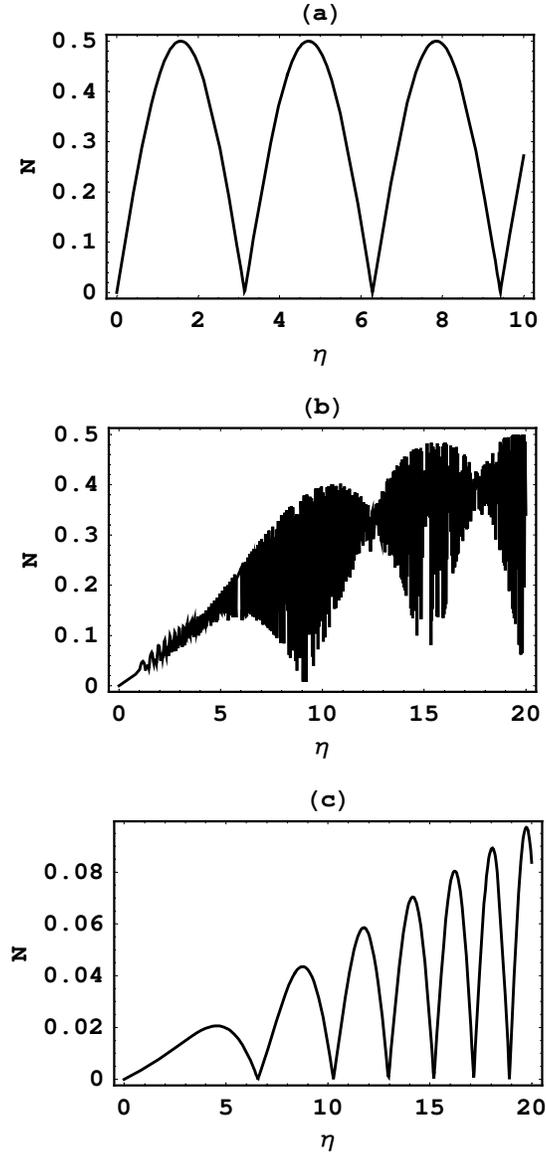} \caption{Negativity as a function of the
coupling constant between the atom and the cavity  $\eta$, when
$T=1$, $\theta=\pi/4$, $\zeta=10$. (a) $n_1=n_2=0$;  (b)
$n_1=n_2=100$; (c) $n_1=0$, $n_2=100$. }
\end{figure}

\begin{figure}
\epsfig{figure=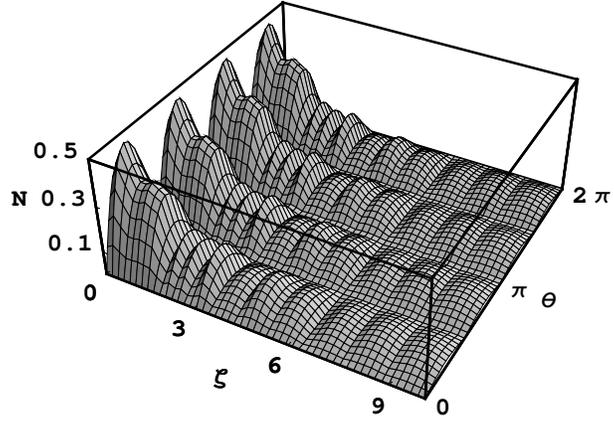} \caption{Surface plot of negativity as a
function of Kerr medium $\zeta$ and phase angle $\theta$, when
$T=1$, $\eta=1$, $n_1=n_2=100$.}
\end{figure}

\begin{figure}
\epsfig{figure=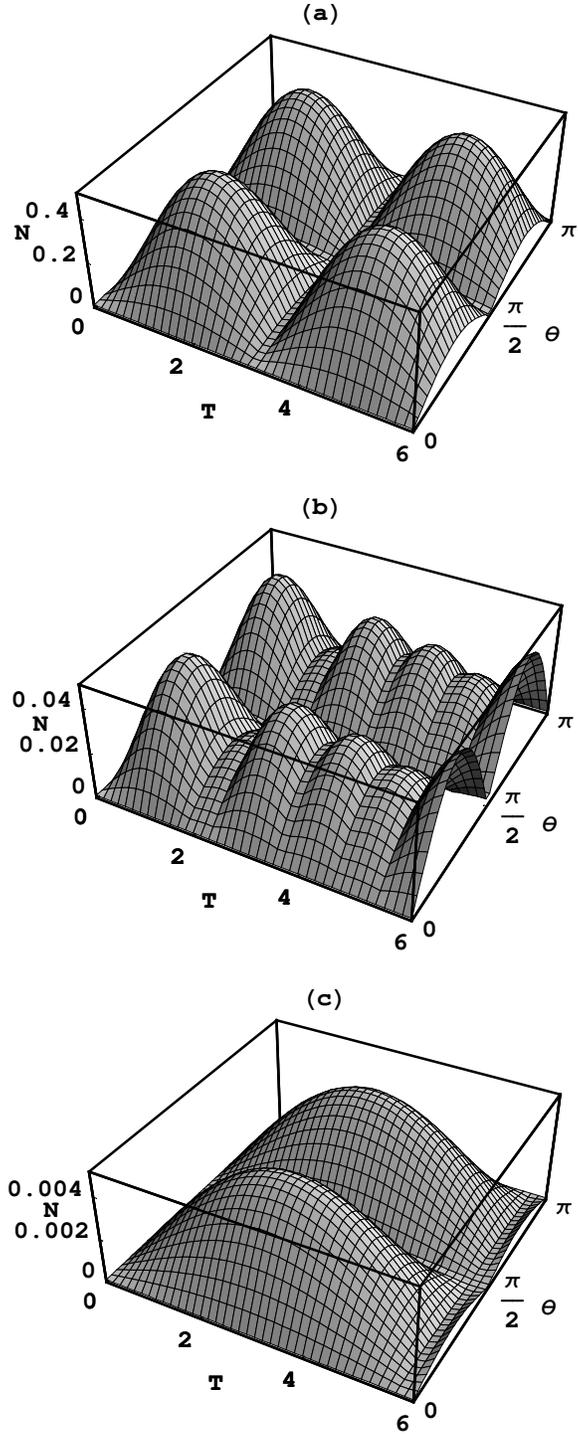} \caption{Surface plot of negativity as a
function of time $T$ and phase angle $\theta$, when $\eta=1$,
$\zeta=10$. (a) $n_1=n_2=0$; (b) $n_1=n_2=100$; (c) $n_1=0$,
$n_2=100$.}
\end{figure}

\end{document}